\documentclass[fleqn,10pt]{wlscirep}
\usepackage[utf8]{inputenc}
\usepackage[T1]{fontenc}

\usepackage{amsmath,amsfonts,amssymb}
\usepackage{siunitx}
\usepackage{enumitem}
\usepackage{graphicx}
\usepackage{caption}
\usepackage{subcaption}
\usepackage{bm}
\usepackage{sectsty}
\usepackage{float}
\usepackage{authblk}
\usepackage{lineno}

\newcommand{\matr}[1]{\mathbf{#1}}

\usepackage{ulem}

\title{A time-consistent stabilized finite element method for fluids with applications to hemodynamics}

\author[1]{Dongjie Jia}
\author[1,*]{Mahdi Esmaily}
\affil[1]{Cornell University, Sibley School of Mechanical and Aerospace Engineering, Ithaca NY, 14850, USA}

\affil[*]{Correspondence: me399@cornell.edu}

\keywords{stabilized finite element method, cardiovascular simulation, computational fluid dynamics}

\begin{abstract}
    Several finite element methods for simulating incompressible flows rely on the streamline upwind Petrov-Galerkin stabilization (SUPG) term, which is weighted by $\tau_{\mathrm{SUPG}}$. 
    The conventional formulation of $\tau_{\mathrm{SUPG}}$ includes a constant that depends on the time step size, producing an overall method that becomes exceedingly less accurate as the time step size approaches zero.
    In practice, such {method} inconsistency introduces significant error in the solution, especially in cardiovascular simulations, where small time step sizes may be required to resolve multiple scales of the blood flow.
    To overcome this issue, we propose a consistent method that is based on a new definition of $\tau_{\mathrm{SUPG}}$. 
    This method, which can be easily implemented on top of an existing streamline upwind Petrov-Galerkin and pressure stabilizing Petrov-Galerkin method, involves the replacement of the time step size in $\tau_{\mathrm{SUPG}}$ with a physical time scale. 
    This time scale is calculated in a simple operation once every time step for the entire computational domain from the ratio of the $L$\textsuperscript{2}-norm of the acceleration and the velocity. 
    The proposed method is compared against the conventional method using {four cases: a steady pipe flow, a blood flow through vascular anatomy, an external flow over a square obstacle, and a fluid-structure interaction case involving an oscillatory flexible beam}. 
    These numerical experiments, which are performed using linear interpolation functions, show that the proposed formulation eliminates the inconsistency issue associated with the conventional formulation in all cases. 
    While the proposed method is slightly more costly than the conventional method, it significantly reduces the error, particularly at small time step sizes. 
    For the pipe flow where an exact solution is available, we show the conventional method can over-predict the pressure drop by a factor of three.
    This large error is almost completely eliminated by the proposed formulation, dropping to approximately 1\% for all time step sizes and Reynolds numbers considered.

\end{abstract}
\begin{document}

\flushbottom
\maketitle
%
%
\thispagestyle{empty}

\section*{Introduction}

Studies of the human cardiovascular system have greatly benefited from the advances in computational fluid dynamics (CFD) since the end of the 20th century\cite{taylor1998finite,taylor1998finite2,taylor1999predictive}. 
Among the numerical methods for solving the Navier-Stokes equations\cite{eymard2000finite,leveque2007finite,hughes2012finite}, the finite element method has found popularity for cardiovascular CFD simulations because it is a convenient framework for dealing with complex geometries and modeling fluid-structure interaction\cite{van2003finite,bazilevs2006isogeometric,bazilevs2009computational,quarteroni2016geometric}.
Together with advancements in clinical imaging techniques\cite{markl2003time,dyverfeldt20154d,schulz2020standardized,markl20124d}, CFD simulations using the finite element method have taken a significant part in in-vitro studies, clinical diagnosis, and surgical planning for cardiovascular diseases\cite{antiga2008image,taylor2009patient,kim2010patient,sankaran2012patient,taylor2013computational,mittal2016computational}.

The finite element method for solving the unsteady Navier-Stokes equations relies on an upwind term that adds an artificial diffusion along the stream-wise directions, weighted by a stabilization parameter $\tau$, to prevent nonphysical oscillations inherent to the Galerkin method in strongly convective regimes\cite{brooks1982streamline,hughes1986new, shakib1991new,franca1992stabilized}.
One of the most commonly used formulations of $\tau$ has been the one proposed in the streamline upwind Petrov-Galerkin formulations (SUPG)\cite{brooks1982streamline}, $\tau_{\mathrm{SUPG}}$, which is also adopted in others such as the residual-based variational multiscale (RBVMS) formulation\cite{akkerman2008role,bazilevs2010large}.

Traditionally, the steady form of $\tau_{\mathrm{SUPG}}$ is derived from a 1D steady advection-diffusion model problem, such that the added diffusion to the Galerkin's formulation is just enough to recover the exact solution, thereby eliminating numerical oscillations at higher element Peclet numbers\cite{brooks1982streamline, hughes1986new,tezduyar1991stabilized,shakib1989finite}.
While this steady form of $\tau_{\mathrm{SUPG}}$ works well for strongly advective steady-state flows, it is not readily applicable to time-varying flows, as it exhibits poor convergence behavior, particularly at smaller time step sizes ($\Delta t$).
The traditional strategy to overcome this issue has been adding a $\Delta t$-dependent term to the definition of $\tau_{\mathrm{SUPG}}$\cite{hughes1986new,shakib1989finite,1998codina}.
The added $\Delta t$-dependent term, which is based on the discrete approximation of the inverse of the strong differential operator, dominates the contributions associated with the advection and diffusion terms at small time step sizes.
This design, which has found widespread use for its excellent convergence characteristics, produces a strong solution dependency on the time step size, such that the solution becomes less accurate as the time step size is reduced toward zero\cite{bochev2004stability,codina2007time,hsu2010improving} (also see the results section).
As a consequence, this time step size-dependent design of $\tau_{\mathrm{SUPG}}$ produces an inconsistent method with regard to the time step size.
This inconsistency issue is particularly exacerbated in a subset of cardiovascular simulations that demand a small time step size for their multiscale behaviors, such as those involving lumped parameter network modeling\cite{quarteroni2003analysis,vignon2006outflow,moghadam2013modular,arbia2014numerical}.
In fact, some of the most popular packages for cardiovascular simulation\cite{updegrove2017simvascular,arthurs2021crimson} are based on a stabilized finite element formulation that also suffers from the described issue, thereby motivating the present study.

There has been some effort in the past to overcome the inconsistency issue associated with this design of $\tau_{\mathrm{SUPG}}$\cite{bochev2004stability, john2008finite,burman2010consistent}.
T.E. Tezduyar and others introduced an element-vector-based $\tau_{\mathrm{SUPG}}$ that uses the relative elemental magnitude of terms in the weak form of the Navier-Stokes equations\cite{tezduyar2000finite,hsu2010improving}.
Although time step size dependency is not entirely eliminated with the proposed formulation, a notable reduction in the solution variation with the time step size was reported in turbulent channel flow simulations. 
Furthermore, this method relies on sets of integrals on each element that must be performed before the evaluation of the discrete form at the Gauss quadrature. 
Thus, it may not be simple to implement this technique in an existing finite element program that is not already designed based on the element-vector-based method. 
In another study, R. Codina and others proposed a subscale-tracking approach that solves a time-dependent ordinary differential equation at each Gauss integration point to evolve its stabilization parameter in time\cite{codina2007time}.
This method eliminates the time step size dependency for steady-state solutions.
The additional computational cost to solve an ordinary differential equation at each Gauss point, however, is non-negligible for this method.
Despite the relatively better time step size consistency shown by these methods in the numerical experiments, they have not found widespread implementation in cardiovascular simulations~\cite{kamensky2015immersogeometric,updegrove2017simvascular}.

In this study, we propose a formulation for $\tau_{\mathrm{SUPG}}$ that eliminates the solution's dependency on the time step size.
This method is simple to implement in existing CFD solvers that are based on the streamline upwind Petrov-Galerkin and pressure stabilizing Petrov-Galerkin (SUPG/PSPG) method.
More specifically, we replace the inverse of the time step size in the definition of the $\tau_{\mathrm{SUPG}}$ with a measure of the flow frequency. 
The motivation behind such formulation lies in the spectral formulation of the unsteady Stokes equations where the time step size dependent parameter in the $\tau_{\mathrm{SUPG}}$ is replaced by the spectral mode number\cite{esmaily2022stabilized}.
The same parameter, when expressed in a spatio-temporal setting, inspires the use of a flow-dependent time scale in the $\tau_{\mathrm{SUPG}}$ definition, hence motivating the present design.

The idea of replacing the time step size in $\tau_{\mathrm{SUPG}}$ with an acceleration-to-velocity ratio has been proposed independently earlier by J. Evans and others\cite{evans2018residual} to simulate turbulent flows.
Our present study nonetheless distinguishes from the previous study both in the formulation and the application. 
First, the previous study adopts an elemental measure of velocity and acceleration, while we use the global-averaged values in defining $\tau_{\mathrm{SUPG}}$. 
As we discuss later in the formulation section, this choice was made for stability reasons. 
While a local measure produces stable results when used in conjunction with an explicit time integration, it creates instabilities in implicit solvers, which is the concern of the present study. 
Second, the present study investigates the behavior of the new $\tau_{\mathrm{SUPG}}$ using a range of canonical and realistic anatomical cases, thereby evaluating its potential for cardiovascular simulations. 

The article is organized as follows: We first present the formulation of the stabilized finite element method for the Navier-Stokes equations, the motivation behind the proposed $\tau_{\mathrm{SUPG}}$, and its formulation.
We present {four} cases to compare the present formulation against the conventional one: a pipe flow with steady boundary conditions, a time-periodic blood flow in a complex cardiovascular geometry (modified Blalock-Taussig shunt), a two-dimensional external flow over a square, {and a two-dimensional flow over a square with an attached flexing beam}.
We will also discuss the convergence and computational cost of the present formulation.
Lastly, we conclude our study and discuss the future outlook in the broader fields of cardiovascular simulations and stabilized finite element methods.

\section*{Formulation}
\label{sec2}

The Navier-Stokes equations for incompressible flows are stated as
\begin{equation}
\begin{split}
    \matr{R_{\mathrm{M}}} &= \rho\left(\frac{\partial \matr{u}}{\partial t} + \matr{u}\cdot\boldsymbol\nabla\matr{u} - \matr{f}\right)-\boldsymbol\nabla\cdot\boldsymbol\sigma &= \matr{0} \qquad \text{in }\Omega \times {\left(0,T\right]},\\
    R_{\mathrm{C}} &= \boldsymbol\nabla\cdot\matr{u} &= 0 \qquad \text{in }\Omega \times {\left(0,T\right]},
\end{split}
\label{strong-NS}
\end{equation}
where $\rho$ is the density, $\matr{u}(\matr{x},t)$ is the velocity, $\matr{f}(\matr{x},t)$ is the external forcing, $\Omega\times [0,T]$ is the fluid computational spatio-temporal domain, and the stress tensor
\begin{equation}
\begin{split}
    \boldsymbol\sigma(p,\matr{u}) &= -p\matr{I}+2\mu\boldsymbol\epsilon(\matr{u}), \\
    \boldsymbol\epsilon(\matr{u}) &= \frac{1}{2}\left((\boldsymbol\nabla\matr{u})+(\boldsymbol\nabla\matr{u})^\intercal\right),
\end{split}
\end{equation}
where $p(\matr{x},t)$ is pressure and $\mu$ is the dynamic viscosity. 
The Dirichlet and Neumann boundary conditions are defined as 
\begin{equation}
\begin{split}
        \matr{u} &= \matr{g} \;\; \mathrm{on \; \Gamma_g}, \\
        \matr{\sigma}\matr{n} &= \matr{h} \;\; \mathrm{on \; \Gamma_h},
\end{split}
\end{equation}
respectively, where $\mathrm{\Gamma_g}$ and $\mathrm{\Gamma_h}$ are subsets of the boundary $\Gamma$ where the Dirichlet and Neumann boundaries are prescribed, $\matr{n}$ is the outward unit normal vector, and $\matr{g}$ and $\matr{h}$ are the given Dirichlet and Neumann boundary conditions, respectively. 

The discrete form of the Navier-Stokes equations we used in this study is stated as finding $\matr{u}^h \in S_\matr{u}^h$ and $p^h \in S_p^h$ such that for all $\matr{w}^h \in V_\matr{u}^h$ and $q^h \in V_p^h$,
\begin{multline}
    \int_\Omega \matr{w}^h\cdot\rho\left(\frac{\partial\matr{u}^h}{\partial t}+\matr{u}^h\cdot\boldsymbol\nabla\matr{u}^h-\matr{f}\right)d\Omega + \int_\Omega\boldsymbol\epsilon\left(\matr{w}^h\right):\boldsymbol\sigma\left(p^h,\matr{u}^h\right)d\Omega -\int_{\Gamma^h}\matr{w}^h\cdot\matr{h}^h d\Gamma +\int_\Omega q^h\boldsymbol\nabla\cdot\matr{u}^h d\Omega \\ 
    +\sum_{e=1}^{n_{el}}\int_{\Omega^e}\left[\tau_{\mathrm{SUPG}}\left(\matr{u}^h\cdot\boldsymbol\nabla\matr{w}^h+\frac{1}{\rho}\boldsymbol\nabla q^h\right)\cdot \matr{R_{\mathrm{M}}}^h + \rho \nu_{\mathrm{C}} \nabla\cdot \matr{w}^h R_{\mathrm{C}}^h\right] d\Omega= 0.
    \label{weak-NS}
\end{multline}
In the above problem statement, $S_\matr{u}^h$ and $S_p^h$ are the discrete solution spaces for the velocity and pressure, respectively, and $V_\matr{u}^h$ and $V_p^h$ are the finite-dimensional test function spaces for the velocity and pressure, respectively. 

In Equation \eqref{weak-NS}, the terms directly obtained from Equation \eqref{strong-NS} are supplemented with three elemental stabilization terms.
The two terms multiplied by $\tau_{\mathrm{SUPG}}$ are the conventional SUPG and PSPG stabilizations to ensure stability in strongly convective flows and allow for equal order interpolation functions for velocity and pressure, respectively\cite{brooks1982streamline,hughes1986new}.
The term involving $\nu_{\mathrm{C}}$ comes from the residual-based variational multiscale methods (VMS)\cite{bazilevs2007variational,bazilevs2008isogeometric,ahmed2017review}.
While there are some variations in defining these terms, we consider
\begin{equation}
    \tau_{\mathrm{SUPG}} = \left(\left(\frac{2}{\Delta t}\right)^2+\matr{u}^h \cdot \boldsymbol{\xi} \matr{u}^h+ C_{\mathrm{I}}\nu^2\boldsymbol{\xi}:\boldsymbol{\xi}\right)^{-\frac{1}{2}},    
\label{eqn1}
\end{equation}
and
\begin{equation}
    \nu_{\mathrm{C}} = \left( \text{tr} \left(\boldsymbol{\xi} \right) \, \tau_{\mathrm{SUPG}} \right)^{-1},
    \label{nuC}
\end{equation}
as the common conventional definition of these stabilization parameters for later comparisons\cite{shakib1989finite, bazilevs2007variational,tezduyar2003stabilization}. 
 In Equation \eqref{eqn1}, $\nu=\mu/\rho$ is the kinematic viscosity, $\Delta{t}$ is the time step size, $\boldsymbol{\xi}$ is the covariant tensor obtained from the mapping of the physical-parent elements, and $C_{\mathrm{I}}$ is a shape-function-dependent constant, which is 3 in our study. 

The inconsistency of the above-mentioned stabilized formulation is caused by $\Delta t$ in $\tau_{\mathrm{SUPG}}$. 
In a steady flow, in which the solution should be independent of $\Delta t$, $\tau_{\mathrm{SUPG}}$ and thus the overall added diffusion will change with $\Delta t$, creating a time step size dependent solution.
In the later sections, we will demonstrate that this inconsistency issue is not unique to steady-state flows and also occurs for unsteady flows. 

In an earlier study, we introduced a pressure-stabilized technique for solving the unsteady Stokes equations expressed in the frequency domain rather than the time domain\cite{esmaily2022stabilized}.
The resulting complex-valued stabilization parameter was derived systematically by taking the divergence of the momentum equation and estimating the Laplacian in the diffusion term using a characteristic element size. 
The modulus of that PSPG-type stabilization parameter is
\begin{equation}
    \left|\tau\right| \propto  \left(\omega^2 + \nu^2\boldsymbol{\xi}:\boldsymbol{\xi}\right)^{-1/2},
    \label{eqn:tau1}
\end{equation}
where $\omega$ is the spectral mode appearing as a source term in the frequency formulation of the unsteady Stokes equations.
This spectral formulation of $\tau$ closely resembles the conventional definition of $\tau_{\mathrm{SUPG}}$ in Equation \eqref{eqn1} if $2/\Delta{t}$ is replaced by $\omega$.
The $\matr{u}^h\cdot\boldsymbol{\xi}\matr{u}^h$ term does not appear in Equation \eqref{eqn:tau1} as the convective acceleration term is not present in the unsteady Stokes equations.
Adding this term into Equation \eqref{eqn:tau1} and incorporating the $O(1)$ constant $C_{\mathrm{I}}$ results in 
\begin{equation}
    \tau_{\mathrm{SUPG}} =  \left(\omega^2+\matr{u}^h \cdot \boldsymbol{\xi} \matr{u}^h + C_{\mathrm{I}} \nu^2\boldsymbol{\xi}:\boldsymbol{\xi}\right)^{-1/2},
    \label{eqn:tauom}
\end{equation}
which is adopted in this study to replace the conventional definition of $\tau_{\mathrm{SUPG}}$ from Equation \eqref{eqn1} when solving Equation \eqref{weak-NS}.

The new definition of $\tau_{\mathrm{SUPG}}$ in Equation \eqref{eqn:tauom} becomes identical to the traditional formulation (Equation \eqref{eqn1}) if $\omega = 2/\Delta{t}$.
This value is close to the largest frequency associated with the time discretization, namely $\pi/\Delta{t}$ that occurs when the solution oscillates between consecutive time steps.
In practice, especially in cardiovascular simulations, the solution is a much smoother function of time and has a frequency content that peaks at a much smaller $\omega$ than $\pi/\Delta{t}$.
The distinction of these two frequencies inspires the proposed definition of $\tau_{\mathrm{SUPG}}$.

It is straightforward to evaluate Equation \eqref{eqn:tauom} in a spectral formulation as $\omega$ is the computed frequency and readily available as an independent parameter. 
However, its adoption is not straightforward in a traditional time formulation, where $\omega$ does not appear as an independent parameter. 
Ideally, $\omega$ must satisfy several properties. 
First, it must produce a scheme that remains stable under a variety of conditions.
Second, it must be extracted from physical variables, such as the velocity and acceleration, rather than the time step size, so that $\tau_{\mathrm{SUPG}}$ converges to a unique quantity as the time step size goes to zero. 
Third, it should be simple to implement and cost-efficient to calculate. 
Given these criteria, we propose
\begin{equation}
    \omega = \frac{\lVert\frac{\partial{\matr{u}^h}}{\partial t}\rVert_{L^2}}{\lVert\matr{u}^h\rVert_{L^2}},
    \label{eqn2}
\end{equation}
where $\lVert \matr{f}\rVert^2_{L^2} = \int_\Omega \lVert\matr{f}\rVert^2 d\Omega$.
This formulation of $\omega$ is designed to go to zero as the flow reaches a steady state, where $\frac{\partial{\matr{u}^h}}{\partial t}$ goes to zero. 
We will show in the results section that this formulation is consistent in both steady and unsteady flows as $\Delta t \to 0$.
We will also show that the present formulation is relatively robust even though it increases the computational cost compared to the conventional method. 

For moving domain simulations that express Equation \eqref{weak-NS} in an arbitrary Eulerian-Lagrangian framework, $\matr{u}^h$ in the convective acceleration term is replaced by the fluid velocity relative to moving mesh $\matr{u}^h - \hat{\matr{u}}^h$, where $\hat{\matr{u}}^h$ denotes mesh velocity. 
It is this velocity that is employed in the definition of $\tau_{\rm SUPG}$ and also $\omega$ in Equation \eqref{eqn2}, changing it to 
\begin{equation}
    \omega = \frac{\lVert\frac{\partial{\matr{u}^h}}{\partial t}|_{\hat{\matr{x}}}\rVert_{L^2_{\Omega^f}}}{\lVert\matr{u}^h - \hat{\matr{u}}^h\rVert_{L^2_{\Omega^f}}},
    \label{omega-ale}
\end{equation}
where the acceleration term is measured at the mesh node location $\hat{\matr{x}}$ and integrals performed over the fluid domain $\Omega^f$. 
By subtracting the mesh velocity from the fluid velocity in Equation \eqref{omega-ale}, the resulting definition of $\omega$ will be Galilean invariant. 
This results in a scheme that produces a unique solution if all velocities were to be measured from a moving inertial reference frame. 
{Later in the result section, we demonstrate the consistency of this method in a moving domain configuration using a fluid structure interaction simulation test case.}

Ideally, one would compute velocity and acceleration locally at the Gauss quadrature point when evaluating $\omega$ so that the solution becomes a function of the local dynamics of the problem. 
Unfortunately, this choice, which has been successfully employed with explicit time integration in the past\cite{evans2018residual}, fails to converge in our implicit formulation. 
This lack of convergence can be attributed to the velocity appearing in the denominator of Equation \eqref{eqn2}, thereby creating widely varying $\omega$ in regions where flow is temporarily stagnant. 
This convergence issue is avoided for all cases tested here by using a global measure of velocity and acceleration through integrating their norms over the entire domain as in Equation \eqref{eqn2}.

In theory, using a global measure of velocity and acceleration could produce a solution that depends on the domain size. 
Consider an external flow over an obstacle in which the $\lVert\frac{\partial{\matr{u}^h}}{\partial t}\rVert_{L^2}$ term receives non-zero contribution only from regions near the obstacle whereas $\lVert{\matr{u}^h}\rVert_{L^2}$ receives a contribution from the entire domain. 
In this setting, $\omega$ goes to zero as the size of the computational domain grows, resulting in a domain size-dependent value.

As we will show later in the results section, this domain-size dependency issue does not translate to inconsistency of the formulation in practice. 
That is because the $\omega^2$ term in Equation \eqref{eqn:tauom} is much smaller relative to the sum of the other two terms. 
In fact, we show that the solution obtained from the proposed formulation is very similar to that of the conventional formulation with a very large $\Delta t$. 
Therefore, increasing the domain size will decrease a parameter in the definition of $\tau_{\mathrm{SUPG}}$ that is already small, thus hardly changing the solution.   

Even though the $\omega^2$ value in $\tau_{\mathrm{SUPG}}$ is very small, dropping it from its definition will create convergence issues, as it has been established in the past\cite{tezduyar1991stabilized,shakib1989finite}. 
The reason that the inclusion of the $\omega^2$ term in $\tau_{\mathrm{SUPG}}$ prevents such scenarios is that a widely varying solution in time will lead to a relatively large $\omega$. 
As a result, the contribution of $\omega$ grows as the solution becomes more unstable, thereby creating a recovery effect that stabilizes the simulation.   

In CFD applications, the velocity field may be initialized from zero, thus creating a divide-by-zero operation in the code when evaluating Equation \eqref{eqn2}. 
To avoid such possibilities, we set $\omega = 2/\Delta{t}$ at the first time step, recovering the conventional definition of $\tau_{\text{SUPG}}$.

As detailed in a previous publication\cite{esmaily2015bi}, we use the implicit generalized-$\alpha$ time integration scheme\cite{Jansen2000305} in our solver. 
The use of this time integration scheme significantly simplifies the implementation of the proposed formulation since velocity and acceleration are readily available as discrete state variables. 
Thus, in our implementation, $\frac{\partial{\matr{u}^h}}{\partial t}$ and $\matr{u}^h$ in Equation \eqref{eqn2} are explicitly computed from those variables in a single operation.
As we will demonstrate in the results, $\omega$ is a slowly varying parameter, thus we perform this operation only once in each time step using the solution at the previous time step.
Given that the generalized-$\alpha$ method also provides access to variables at the intermediate time points (i.e., $n+\alpha_m$ and $n+\alpha_f$ for the acceleration and velocity, respectively), one may elect to use those intermediate variables to update $\omega$ within each Newton-Raphson iterations. 
This choice, however, is not adopted in this study as it entails extra computations and has little effect on the overall stability of the solver.
The rest of the implementation, including the computation of $\tau_{\mathrm{SUPG}}$ at the Gauss quadrature points based on the intermediate variables are left unchanged and are identical to the conventional formulation.

\section*{Simulations and Results}
\label{sec3}
The above formulation is implemented in our in-house finite-element solver, multi-physics finite-element solver (MUPFES)\cite{moghadam2013modular,esmaily2015bi,esmaily2013new}.
A specialized iterative algorithm, preconditioner, and parallelization strategy are employed for an efficient and scalable solution of the linear system of equations\cite{esmaily2013new,esmaily2015bi,esmaily2015impact,marsden2015multiscale}.
The solver has been verified\cite{steinman2013variability} and extensively employed for cardiovascular modeling in the past\cite{esmaily2012optimization,esmaily2015assisted,jia2021efficient}.
This solver is parallelized using a message passing interface (MPI).
The workload is parallelized using spatial partitioning by employing ParMETIS library\cite{METIS}. 
All computations are performed on a cluster of AMD Opteron$^{\rm TM}$ 6378 processors that are interconnected via a QDR Infiniband.

At each time step, several Newton-Raphson iterations are performed to ensure the residual falls by over three orders of magnitude. 
At each Newton-Raphson iteration, a linear system is solved using the generalized minimal residual (GMRES) method\cite{saad1986gmres} with a tolerance of $10^{-2}$.

{Four} cases are simulated using both the conventional formulation (Equation \eqref{eqn1}) and the present formulation (Equation \eqref{eqn:tauom}) for $\tau_{\text{SUPG}}$: a pipe flow, flow in a modified Blalock-Taussig shunt geometry\cite{esmaily2012optimization}, a two-dimensional external flow over a square, {and a flow over a square with an attached flexible beam.}
These numerical experiments are designed to stress-test various aspects of the two formulations in canonical and physiologic settings.
More specifically, these cases represent three classes of flow simulations where 1) the boundary conditions and the solution are both steady, 2) the boundary condition and the solution are both unsteady, 3) the boundary conditions are steady but the solution is unsteady (in this case due to vortex shedding), {and 4) the fluid domain is not fix with unsteady solution.}
All of the simulations are initialized using $\matr{u_0} = 0$ and continued in time to reach cycle-to-cycle convergence or steady-state solutions.
For apple-to-apple comparison, all parameters, except for the $\tau_{\text{SUPG}}$ definition, are kept the same when comparing the present formulation against its conventional counterpart.

\subsection*{Steady pipe flow}
\label{sec:pipe}
We first consider the case of flow in a straight pipe with steady boundary conditions, which can be considered the most simple and fundamental flow in the cardiovascular system with an existing analytical solution for comparison.
In our simulations, a steady flow rate is imposed at the inlet and the pressure drop across the pipe is predicted using the present and conventional methods. 
The pipe has a length of 15 cm and a radius of 1 cm. 
A parabolic velocity profile is imposed at the inlet with an amplitude that results in a 10 mL/s flow rate.
A zero Neumann boundary condition is imposed at the outlet. 
The dynamic viscosity is fixed at 1 g/cm-s.
Three different densities, 1.571, 15.71, and 157.1 g/mL are selected to produce three Reynolds numbers (Re), 10, 100, and 1000, respectively. 
This way, we capture a wide range of Reynolds numbers that occur in cardiovascular flows\cite{MILLER2012937}.
For all cases, the pressure drop must remain the same according to the Hagen–Poiseuille analytical solution\cite{sutera1993history}, thereby allowing us to measure the accuracy of each method. 

The mesh generated for this case contains $207,063$ tetrahedral elements, which are used for the velocity and pressure interpolation as well as their test functions. 
Each Reynolds number is simulated using four different time step sizes of $\Delta{t} = 10^{-1}$, $10^{-2}$, $10^{-3}$, and $10^{-4}$ seconds.
This range of time step sizes, which is typical in cardiovascular simulations, results in Courant–Friedrichs–Lewy numbers ($\mathrm{CFL} = \overline{u}\Delta t/\overline{\Delta x}$) ranging from $5.2$ to $5.2\times10^{-3}$, based on the mean element size of the mesh, $\overline{\Delta x}$, and the mean flow velocity, $\overline{u}$. 
This range of CFL numbers, which captures under-resolved to over-resolved time discretizations, can be encountered in a typical simulation due to the differences in velocity and mesh resolution in different cardiovascular branches.
Furthermore, multi-domain simulations can require the use of a smaller time step size (hence CFL) to ensure stability\cite{moghadam2013modular}.

All simulation cases were run in parallel with 32 cores. 
Equations are integrated for five seconds (approximately three flow-through times) to ensure steady conditions are reached. 
The $l^2$-norm of the residual is dropped by 3.5 orders of magnitude at each time step using Newton-Raphson iterations. 
Considering three Reynolds numbers, four time step sizes, and two formulations, we ran a total of 24 simulations for this case. 

\begin{figure} [H]
\centering
  \captionsetup{position=bottom}
  \includegraphics[width=\textwidth]{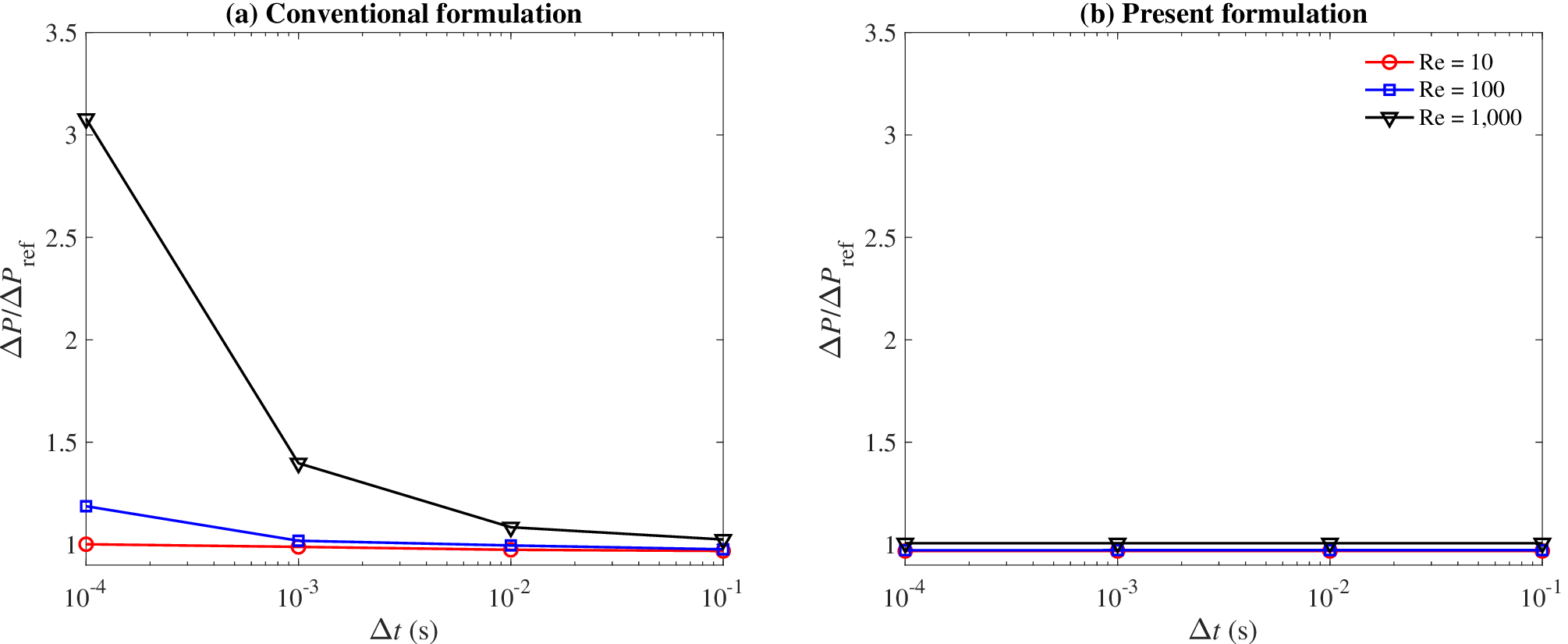}
  \caption{The predicted pressure drop normalized by the analytical solution ($\Delta{P}/\Delta{P}_{\mathrm{ref}}$) for the steady pipe flow case as a function of the time step size ($\Delta{t}$) for (a) the conventional formulation and (b) the present formulation. The three Reynolds numbers are 10 (red circle), 100 (blue square), and 1,000 (black triangle).}
  \label{fig:pipe}
\end{figure}

The results for this case are condensed in Figure~\ref{fig:pipe}, which shows the predicted pressure drop, normalized by that of the Poiseuille solution\cite{sutera1993history}, as a function of the time step size for three different Reynolds numbers.
The solutions calculated using the conventional $\tau_{\mathrm{SUPG}}$ formulation becomes exceedingly less accurate as the Reynolds number is increased and the time step size is reduced (Figure~\ref{fig:pipe}(a)).
That produces a very large error at ${\rm Re} =$1,000 and $\Delta{t} = 10^{-4}$ where the predicted pressure drop is three times that of the analytical prediction.
This large deviation from the reference solution confirms the inconsistency of the conventional formulation that we discussed earlier.
This effect is amplified at higher Reynolds numbers when the artificial viscosity introduced through $\tau_{\mathrm{SUPG}}$ is larger in comparison to the physical viscosity, thus creating a larger variation in the solution as $\tau_{\mathrm{SUPG}}$ is varied with $\Delta t$. 

The present formulation, on the other hand, produces predictions that are independent of the time step size, confirming that it is a consistent formulation for steady-state flows (Figure~\ref{fig:pipe}(b)).
The overall error, which is primarily associated with spatial discretization, is negligible in comparison to the conventional formulation. 
The error for the case discussed above ($\rm Re=1,000$ and $\Delta t=10^{-4}$) drops from $300\%$ to $0.7\%$ when one uses the present rather than the conventional formulation. 

\begin{figure}[H]
    \centering
    \includegraphics[width=0.475\textwidth]{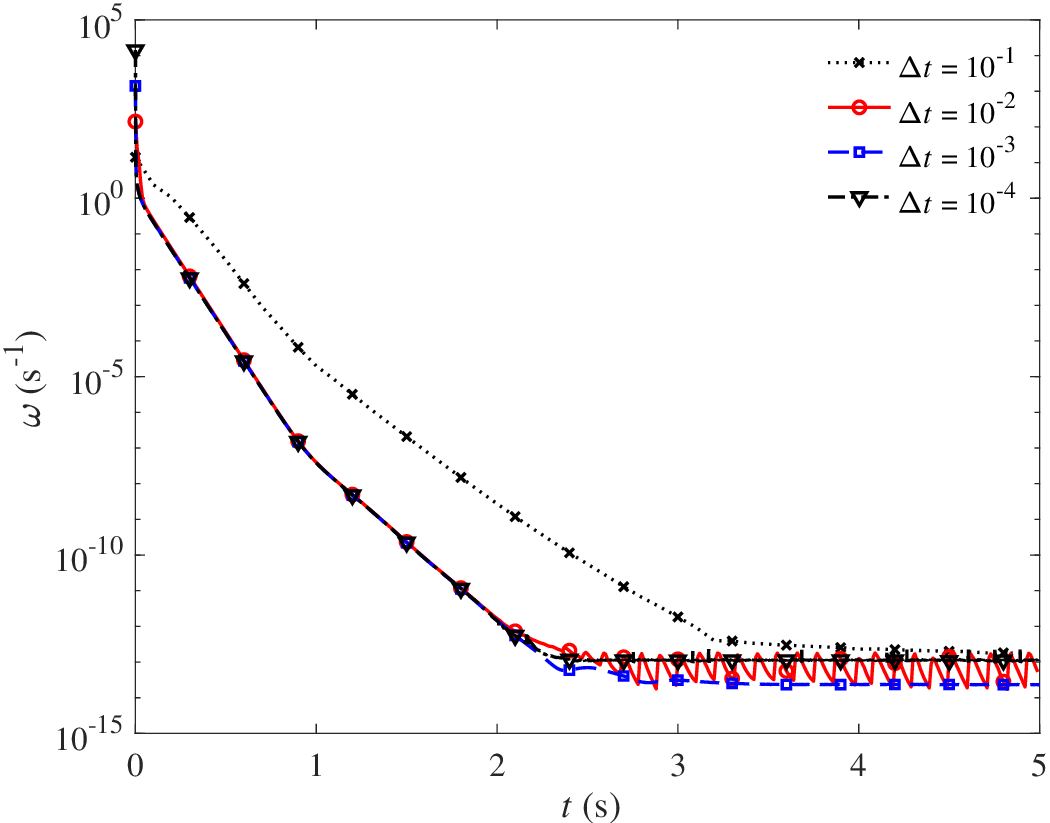}
    \caption{The time evolution of $\omega$ (Equation \eqref{eqn2}) over the course of the simulation for the pipe flow case at Re = 10 and at four different time step sizes of $\Delta{t} = 10^{-1}$ (dotted), $10^{-2}$ (solid), $10^{-3}$ (dashed), and $10^{-4}$ (dot-dashed) seconds.}
    \label{fig:pipeomval}
\end{figure}

Such a large difference in the solutions can be attributed to the large difference between $\omega$ and $2/\Delta t$ term in $\tau_{\text{SUPG}}$ definition. 
The large difference between the two is depicted in Figure~\ref{fig:pipeomval} which shows the history of $\omega$ over the course of the simulation.
Given that $\omega = 2/\Delta t$ at $t=0$, the two methods are equivalent at the beginning of the simulation. 
However, as time progresses, the conventional formulation will significantly deviate from the present formulation by producing an $\omega$ that differs by around fifteen orders of magnitude. 

Given these attractive results and the fact that $\omega \to 0$ in the present formulation, one may be tempted to entirely drop $2/\Delta t$ term from the definition of $\tau_{\mathrm{SUPG}}$.
As we argued earlier, such a modification results in a method that fails to converge particularly at a relatively small time step size and in cases where the flow is highly unsteady. 
Such regime corresponds to the initial stage of the pipe flow simulation, where the flow is rapidly evolving and $\omega \ne 0$ (Figure~\ref{fig:pipeomval}). 
Therefore, incorporating $\omega$ into the definition of $\tau_{\mathrm{SUPG}}$ plays a crucial role in improving the stability of the overall scheme.  

\begin{figure} [H] 
  {\captionsetup{position=bottom}
    \centering
    \includegraphics[width=0.7\textwidth]{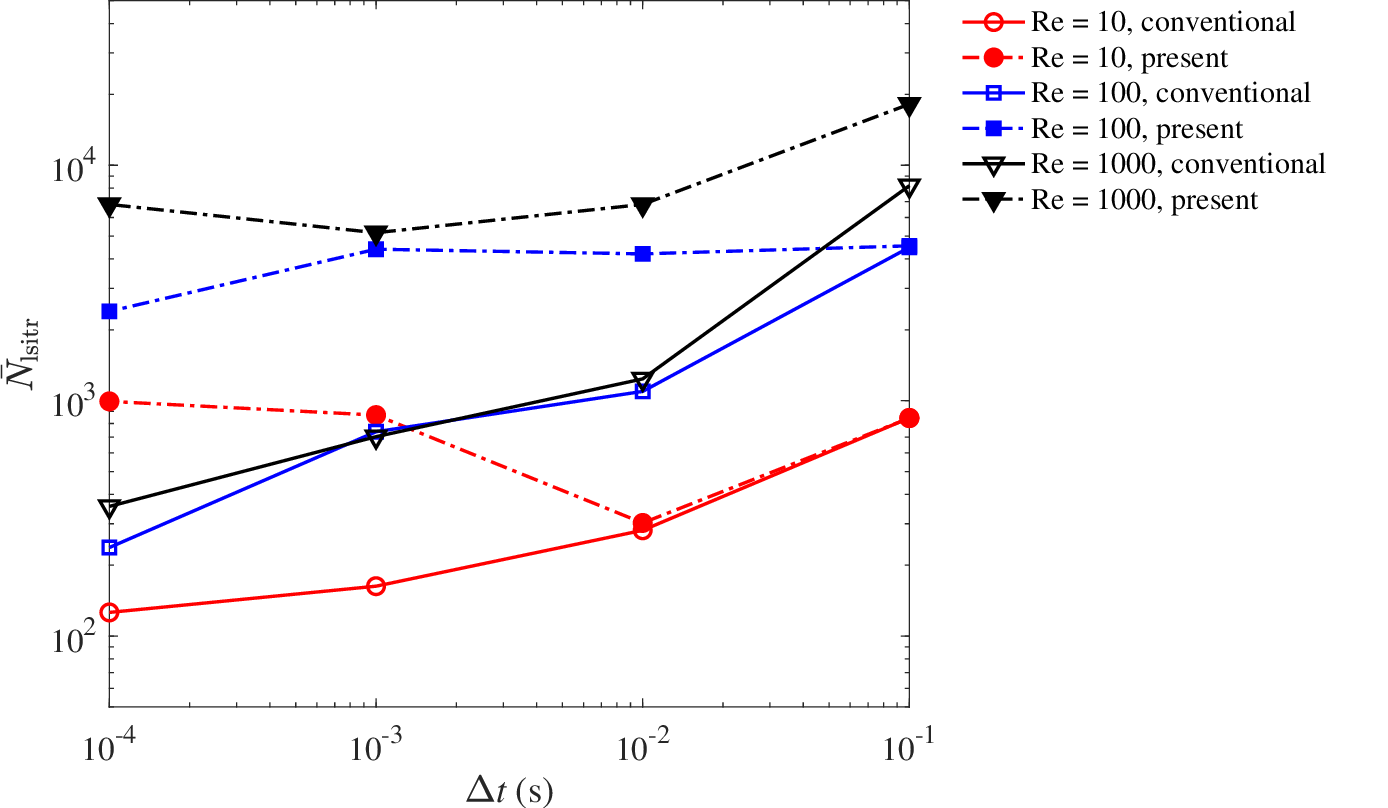}
  \caption{The average number of the linear solver (GMRES) iterations per time step ($\bar{N}_{\mathrm{lsitr}}$) as a function of the time step size ($\Delta{t}$) for the steady pipe flow case. 
  The results correspond to the three simulated Reynolds numbers: 10 (red circle), 100 (blue square), and 1,000 (black triangle) using the conventional (solid line) and the present formulation (dot-dashed line) of $\tau_{\text{SUPG}}$.}
  \label{fig:pipeitr}
}
\end{figure}

Lastly, we must note that although the present method converges for all cases considered, it produces a linear system that is stiffer than that of the conventional formulation. 
As a consequence, the solution of the linear system through an iterative solver will require more iterations, which can be over an order magnitude per time step when compared against the conventional method (Figure~\ref{fig:pipeitr}).
This larger number of iterations translates to a higher overall cost of these calculations, which is on average twice higher than that of the conventional formulation.

\subsection*{Blood flow in vascular anatomy}

\begin{figure}[H]
\centering
  \includegraphics[width=0.9\textwidth]{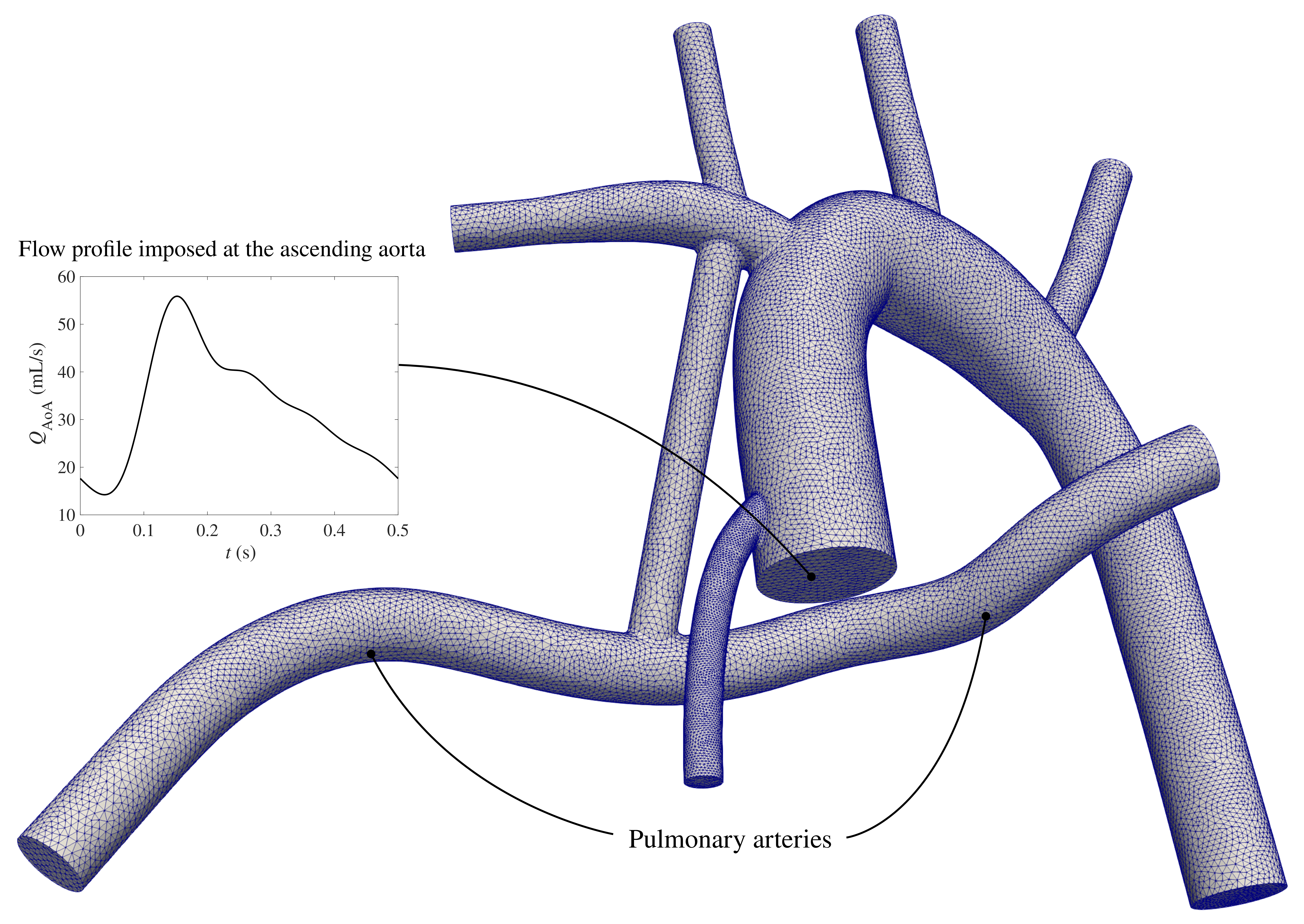}
  \caption{The meshed geometry and the inlet condition (the ascending aorta flow rate, $Q_{\text{AoA}}$) for the modified Blalock-Taussig shunt simulation.}
  \label{fig:btshuntgeo}
\end{figure}

In this case, we compare the performance of the conventional formulation and our present formulation in a realistic cardiovascular geometry with a wide range of Reynolds and CFL numbers. 
The adopted anatomy represents that of an infant who has undergone the modified Blalock-Taussig shunt procedure\cite{esmaily2012optimization,esmaily2015assisted} (Figure~\ref{fig:btshuntgeo}).
The geometry contains multiple branches, among which an unsteady flow with a parabolic profile is imposed at the ascending aorta, which is interpolated from earlier multi-domain simulations\cite{jia2021efficient,jia2022characterization}.
Zero Neumann boundary conditions are imposed on all other branches, which are non-physiological, but nevertheless selected to highlight the difference between the two formulations. 
Since the flow rate through the pulmonary arteries is critical in understanding the performance of the shunt for this procedure, it is used here for simulation results comparison.
The blood is assumed Newtonian with a density of $1.06$ g/mL and a dynamic viscosity of $0.04$ g/cm-s. 
The Reynolds number ranges from 50 to 1300 depending on the branch and the time within the cardiac cycle.
Although the primary source of flow unsteadiness can be traced to the unsteady boundary condition, the complex geometry may also induce more local variations in the solution as a function of time, in this case (namely, generating frequency content in the solution that is not present in the boundary condition). 

We performed simulations using time step sizes of $\Delta{t} = 2.5 \times 10^{-2}$, $2.5 \times 10^{-3}$, and $2.5 \times 10^{-4}$ seconds, which are within the range of actual time step sizes used for these types of studies\cite{esmaily2012optimization,esmaily2015assisted,jia2021efficient}.
The geometry is discretized using $400,936$ tetrahedral elements (Figure~\ref{fig:btshuntgeo}).
All simulations are run in parallel with 288 processors.
Simulations are continued for at least six cardiac cycles (3 seconds) to ensure cycle-to-cycle convergence.

\begin{figure} [ht]
\centering
  \captionsetup{position=bottom}
  \includegraphics[width=\textwidth]{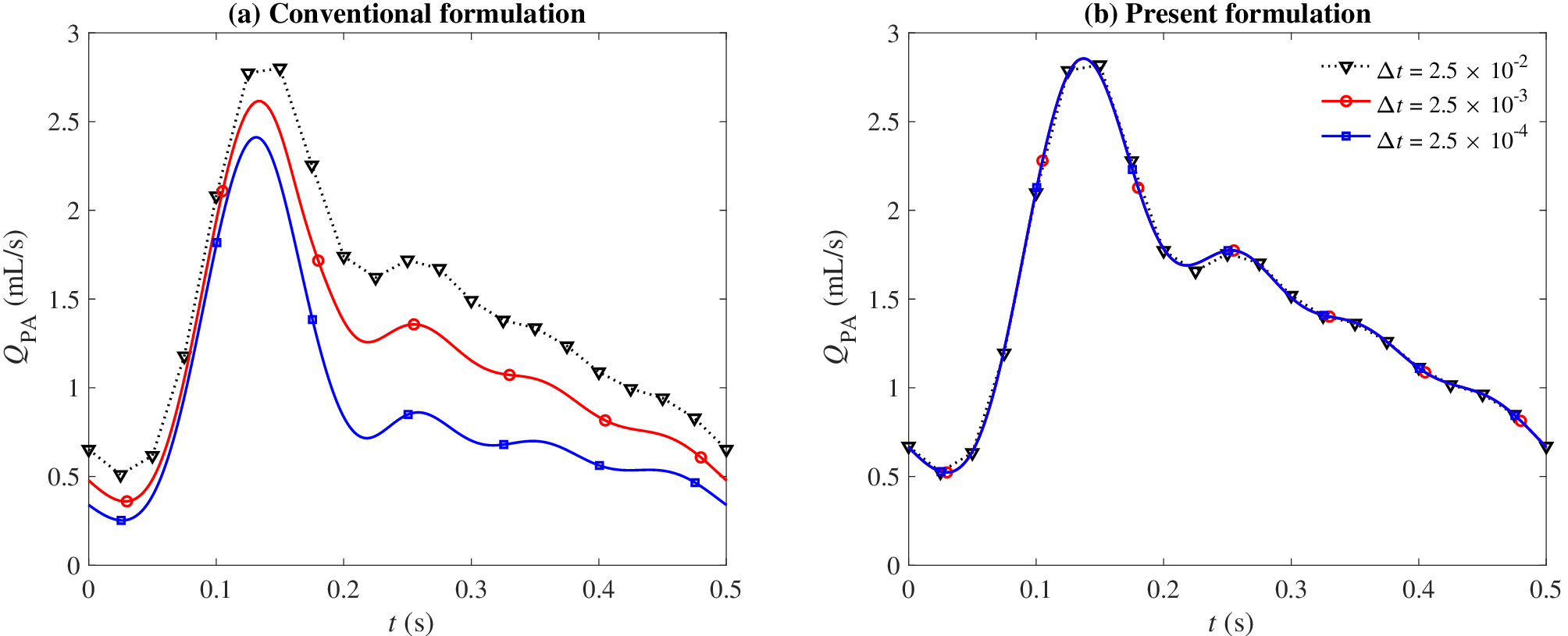}
  \caption{The predicted flow rate through the pulmonary arteries ($Q_{\text{PA}}$) using $\Delta{t} = 2.5 \times 10^{-2}$ (dotted), $2.5 \times 10^{-3}$ (red circle), and $2.5 \times 10^{-4}$ (blue square) seconds, when the computations are performed using (a) the conventional formulation and (b) the present formulation of $\tau_{\text{SUPG}}$. }
  \label{fig:mbts}
\end{figure}

The predicted flow rate through the pulmonary arteries is extracted as a parameter of interest and plotted over one cardiac cycle in Figure~\ref{fig:mbts}. 
For the conventional formulation, we can see a similar trend of solution deterioration as the time step size gets smaller. 
There is a 10\% change in flow rate as the time step size is reduced from $2.5 \times 10^{-2}$ to $2.5 \times 10^{-3}$ seconds. There is another 15\% deviation in the prediction of the conventional formulation when the time step size is further reduced by another order of magnitude, from $2.5 \times 10^{-3}$ to $2.5 \times 10^{-4}$ seconds. 
Such large variations in the results are rather alarming because 500 times steps per cardiac cycle or more (corresponding to $\Delta t\le 10^{-3}$) is a very modest number and has been commonly used in the past cardiovascular CFD studies\cite{esmaily2012optimization, esmaily2015assisted,verma2018optimization,jia2021efficient,shang2019patient,sankaran2012patient}.
According to our numerical experiment with the conventional method, such a commonly used time step size is already too small that it produces substantial error in the results. 

\begin{figure} [ht]
\centering
  \captionsetup{position=bottom}
  \includegraphics[width=0.475\textwidth]{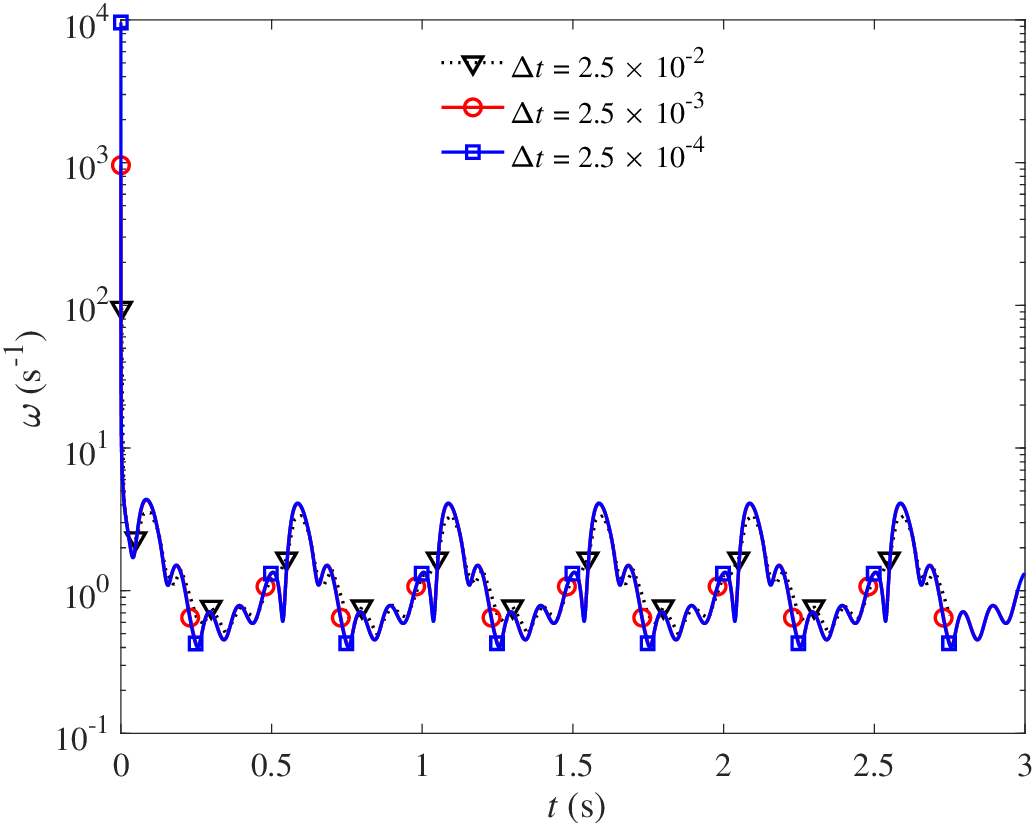}
  \caption{$\omega$ value used in the present formulation as the simulations progress in time for three different time step sizes, $\Delta{t} = 2.5 \times 10^{-2}$ (dotted), $2.5 \times 10^{-3}$ (red circle), and $2.5 \times 10^{-4}$ (blue square) seconds. }
  \label{fig:mbts_om}
\end{figure}

In contrast to the conventional formulation, the method proposed is consistent with results that are almost independent of the time step size (the observed changes were less than 0.1\%). 
$\omega$ values using the present formulation is shown in Figure~\ref{fig:mbts_om} for three different time step sizes. 
At the beginning of the simulations, $\omega$ quickly drops from that of the conventional formulation values to physics-based periodic values. 
As expected, the converged periodic $\omega$ values are almost independent of the time step size, thus resulting in predictions that do not change as $\Delta t \to 0$.

The total CPU time of the computations performed with the present formulation is around 1.5 times that of the conventional formulation, which is acceptable but warrants future improvements.
Simulation results generated with the present formulation will enable researchers and clinicians to have high resolution in terms of time discretization without concerns about solution accuracy.

\subsection*{\label{sec:square}Flow over a square}

In this case, we will present a more textbook unsteady CFD numerical experiment to demonstrate that the consistency issue is not unique to cardiovascular applications.
We will demonstrate that the present formulation solves the inconsistency issue for this case but also demonstrate where the limitation of our formulation currently lies.
We will also use this case to discuss the domain size dependency of (or lack thereof) the present formulation for external flows. 

\begin{figure}[H]
\centering
  \includegraphics[width=0.8\textwidth]{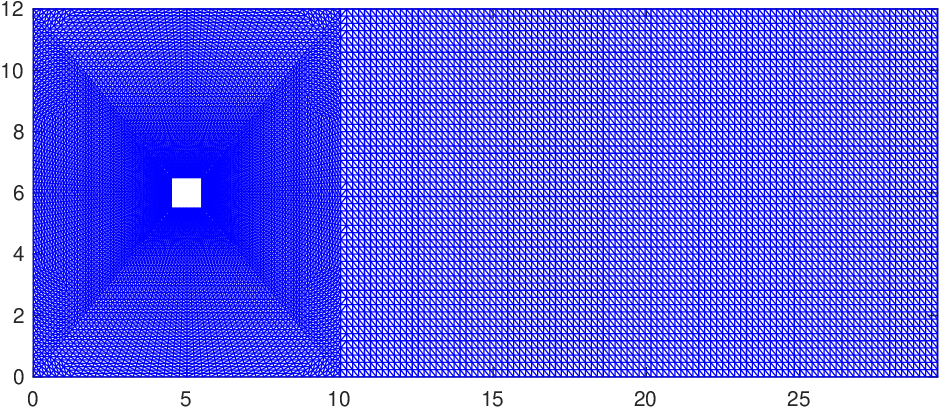}
  \caption{The mesh constructed for the flow over a square obstacle simulation.}
  \label{fig:meshsq}
\end{figure}

We consider a two-dimensional unsteady flow over a square object with a steady inflow boundary condition.
The geometry and mesh used for this case are shown in Figure~\ref{fig:meshsq}.
The square has a side length of 1 m in a 12 m by 29.2 m fluid domain. 
The square obstacle is centered vertically and placed at a distance of 5 m from the inlet on the left side of the domain. 
Uniform horizontal flow with a velocity magnitude of 51.3 m/s is prescribed at the inlet and the outlet is a zero Neumann boundary. 
The top and bottom of the domain are both no-penetration boundaries ($u_y$ = 0) with zero traction in the horizontal direction ($h_x = 0$).
The fluid has a density of $1.18 \times 10^{-3}$ kg/m\textsuperscript{3} and a dynamic viscosity of $1.82 \times 10^{-4}$ kg/m-s.
The Reynolds number of this case is 332, which results in vertex shedding downstream of the obstacle.

The mesh generated for this case contains $28,502$ triangular elements.
Three different time step sizes are considered, $\Delta{t} = 10^{-3}$, $4 \times 10^{-4}$, and $10^{-4}$ seconds. 
At each time step, the time integration residual is dropped by more than three orders of magnitude through Newton-Raphson iterations.
The simulations are continued for 5 seconds to ensure statistically stationary conditions are established.
We used 16 cores to perform these calculations.

\begin{figure} [H]
\centering
  \captionsetup{position=bottom}
  \includegraphics[width=\textwidth]{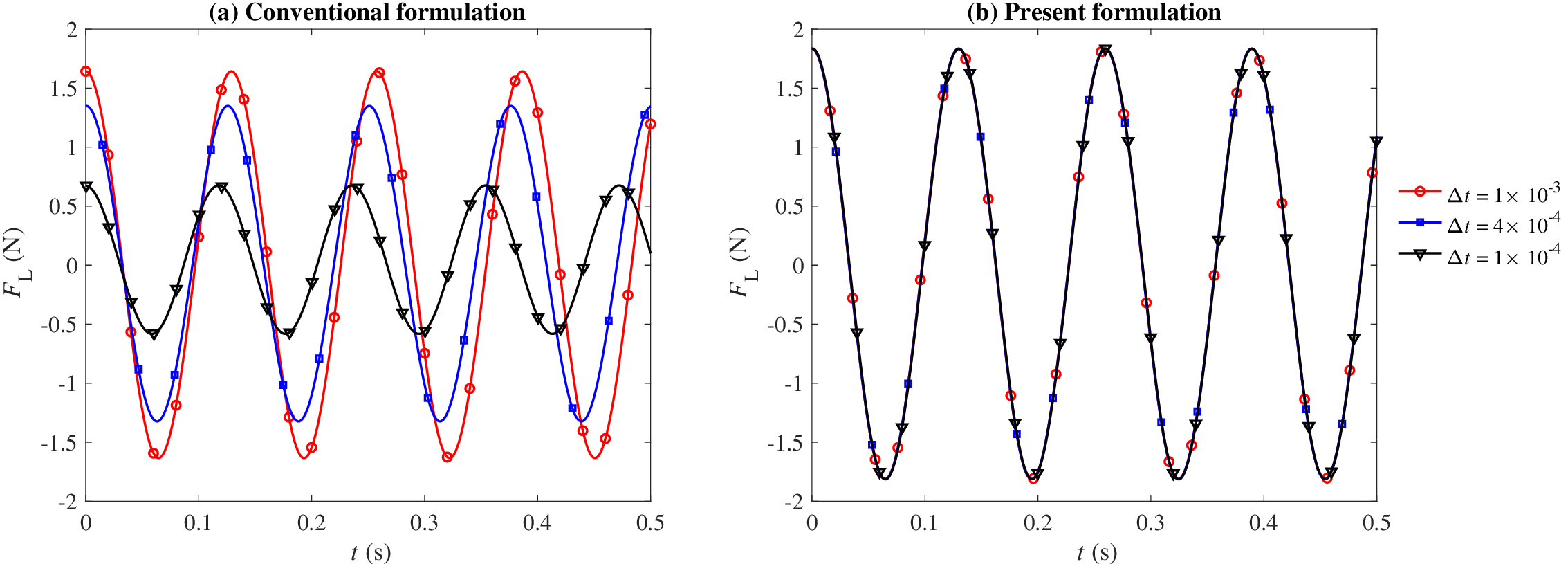}
  \caption{Predicted lift on the obstacle ($F$\textsubscript{L} using $\Delta{t} = 10^{-3}$ (red circle), $4 \times 10^{-4}$ (blue square), and $10^{-4}$ (black triangle) seconds, where $\tau_{\text{SUPG}}$ is computed using (a) the conventional formulation and (b) the present formulation of $\tau_{\text{SUPG}}$. Re = 332.}
  \label{fig:oversq}
\end{figure}

\begin{figure}[H]
    \centering
    \includegraphics[width=\textwidth]{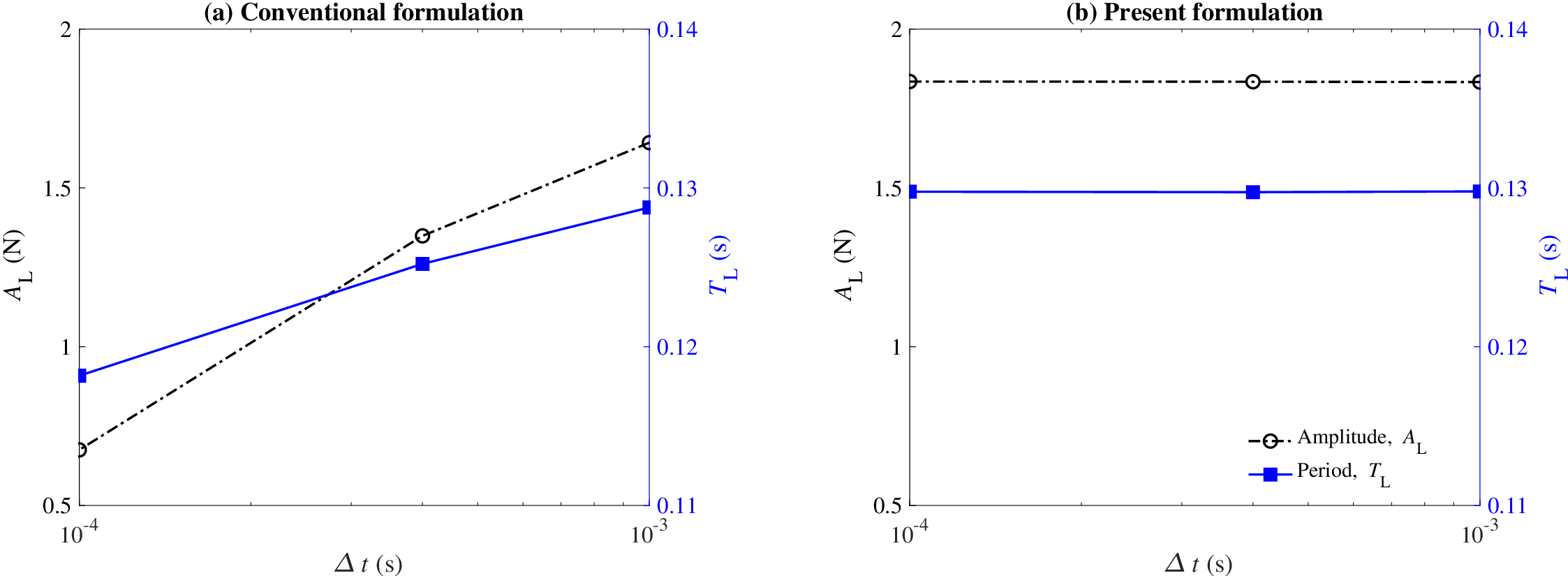}
    \caption{Predicted amplitude (black dash-dotted line) and period (blue solid line) of the lift profile as a function of the time step size ($\Delta{t}$) using (a) the conventional formulation and (b) the present formulation.}
    \label{fig:sqat}
\end{figure}

\begin{figure} [H] 
  {\captionsetup{position=bottom}
    \centering
    \includegraphics[width=\textwidth]{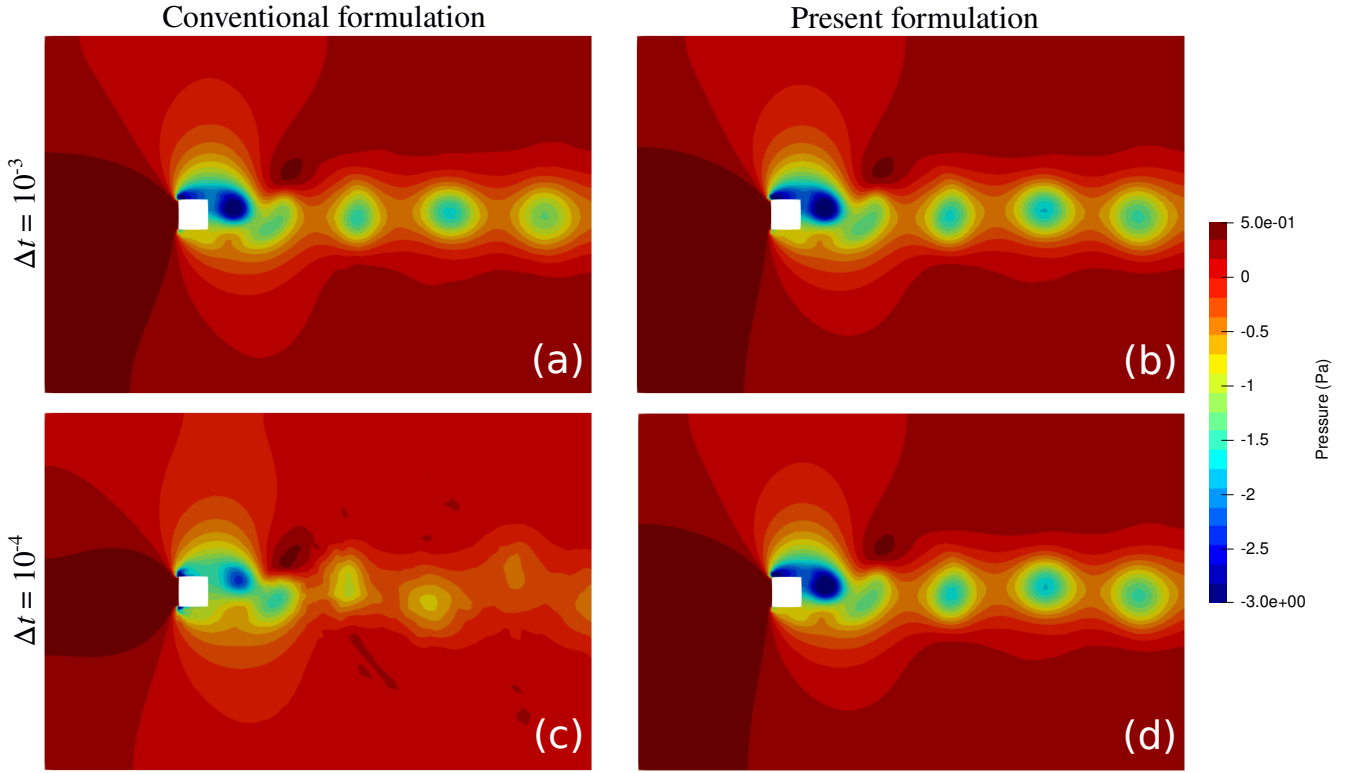}
  \caption{Pressure contour for the flow over a square obstacle (Figure~\ref{fig:meshsq}) captured at the maximum lift. Re = 332. 
  (a) and (c) are the results obtained from the conventional formulation while (b) and (d) are the results obtained from the present formulation. 
  (a) and (b) are obtained using $\Delta{t} = 10^{-3}$ and (c) and (d) using $\Delta{t} = 10^{-4}$.}
  \label{fig:oversqcont}
}
\end{figure}

The lift exerted on the square obstacle during the last 0.5 seconds of the simulation is shown in Figure~\ref{fig:oversq}.
Consistent with what we observed in the previous cases, this case also shows significant improvement in the results when the present design of $\tau_{\text{SUPG}}$ is adopted.
The conventional formulation prediction of the oscillation amplitude and period strongly depends on the time step size, with both values decreasing as $\Delta t \to 0$ (Figure~\ref{fig:sqat}a).
In contrast, little change in these predictions is observed when the presented formulation is adopted (Figure~\ref{fig:sqat}b).
The contrast between the two methods can also be observed when comparing the pressure contours in Figure~\ref{fig:oversqcont}.
The snapshots shown in this figure are taken when the obstacle experiences a maximum lift.
The dependency and lack of dependency of the conventional and present formulation on the time step size are evident in this figure.

The results from the two formulations are compared in terms of mean drag coefficient $\overline{C_\mathrm{d}}$, root mean square of drag coefficient fluctuation $C_\mathrm{d}'$, mean lift coefficient $\overline{C_\mathrm{l}}$, root mean square of lift coefficient fluctuation $C_\mathrm{l}'$, and Strouhal number St in Table~\ref{tab:sqaure}.
This comparison confirms our earlier observation that the present formulation nearly eliminates the dependence of the bulk flow parameters on the time step size.
\begin{table}[H]
\centering
\begin{tabular}{c c c c c c} 
 & $\overline{C_\mathrm{d}}$ & $C_\mathrm{d}'$ & $\overline{C_\mathrm{l}}$ & $C_\mathrm{l}'$ & St\\ 
 \hline
 \multicolumn{1}{l}{\textbf{Conventional}} & & & & & \\
 $\Delta{t} = 1\times 10^{-3}$ & 1.74 & 0.04 & $0.00$& 0.74 & 0.15 \\
 $\Delta{t} = 4\times 10^{-4}$ & 1.81 & 0.03 & $-0.01$& 0.61 & 0.16 \\
 $\Delta{t} = 1\times 10^{-4}$ & 2.07 & 0.01 & $-0.03$& 0.29 & 0.16 \\
 \hline
 \multicolumn{1}{l}{\textbf{Present}} & & & & & \\
 $\Delta{t} = 1\times 10^{-3}$ & 1.72 & 0.04 & $-0.01$ & 0.83 & 0.15 \\
 $\Delta{t} = 4\times 10^{-4}$ & 1.72 & 0.04 & $-0.01$ & 0.83 & 0.15 \\
 $\Delta{t} = 1\times 10^{-4}$ & 1.72 & 0.04 & $-0.01$ & 0.83 & 0.15 \\
 \hline
\end{tabular}
\caption{\label{tab:sqaure} Comparison of the bulk flow parameters for the flow over a square object case between the conventional and present formulations using three different time step sizes. $\mathrm{Re} = 332$.}
\end{table}

Similar to the steady pipe flow, the convergence of the solutions as the time step size decreases is tied to the convergence of $\omega$.
With the present formulation, the value of $\omega$, although not steady, converges to periodic values around 3 regardless of $\Delta t$.
The value of $\omega^2$, which is approximately 9 s$^{-2}$, can be contrasted against $(2/\Delta t)^2$ that ranges from $4\times 10^6$ to $4\times 10^8$ s$^{-2}$ for the simulated cases. 
That large change generates a significant variation in the solution as $\Delta t$ is reduced in the conventional formulation.

\begin{figure} [ht]
\centering
  \captionsetup{position=bottom}
  \includegraphics[width=0.9\textwidth]{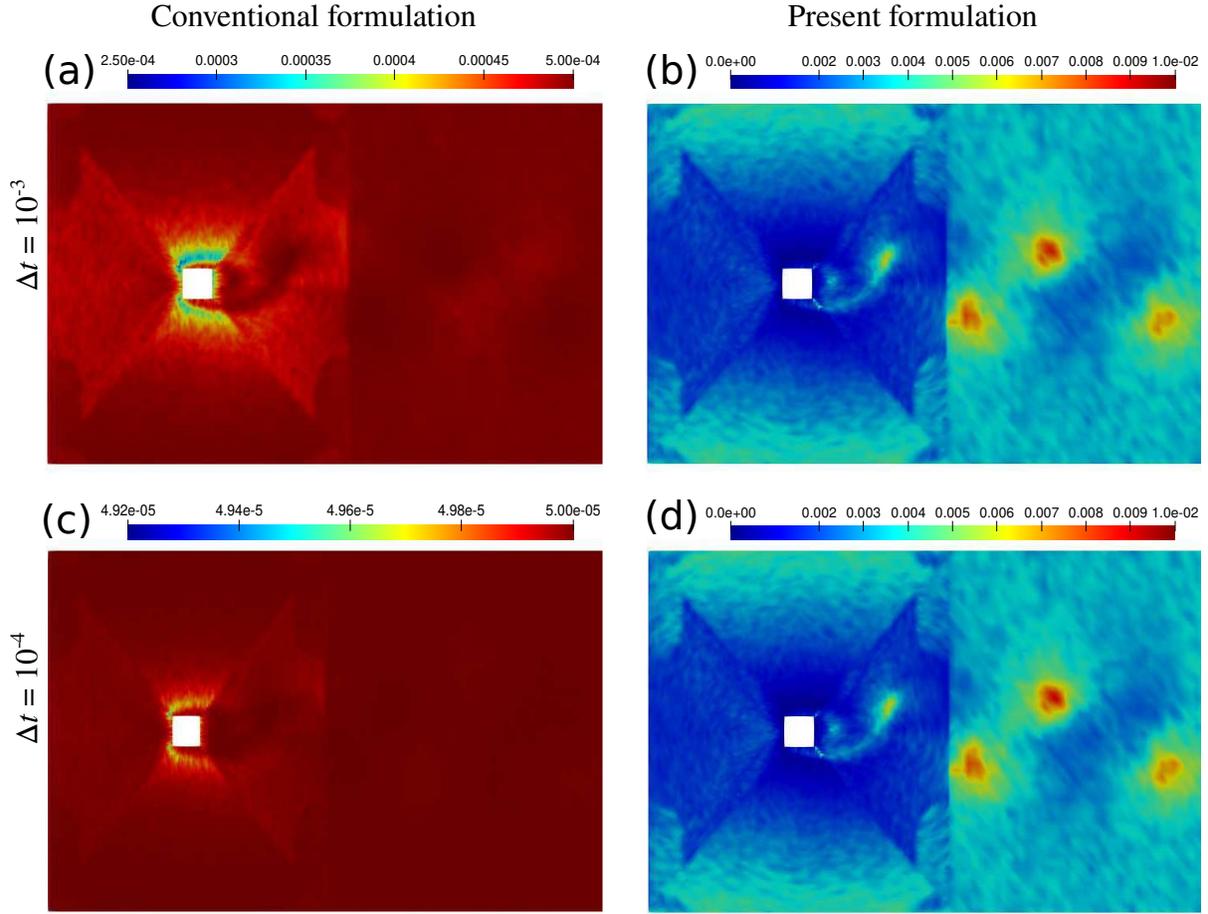}
  \caption{The snapshot of $\tau_{\text{SUPG}}$ obtained at peak lift for the conventional formulation (a,c) and the present formulation (b,d) with $\Delta{t} = 10^{-3}$ (a,b) and $10^{-4}$ (c,d) seconds. Re = 332. Note the range used for the color bars. }
  \label{fig:sq_tau}
\end{figure}

To better see the effect of $2/\Delta t$ term on $\tau_{\text{SUPG}}$, we have produced a snapshot of $\tau_{\text{SUPG}}$ over the entire computational domain for the two formulations in Figure~\ref{fig:sq_tau} at two different time step sizes. 
For the conventional formulation and in particular at $\Delta t= 10^{-4}$, $\tau_{\text{SUPG}}$ is very small and approximately equal to $\Delta t/2$ over the entire domain. 
In the vicinity of the obstacle, where flow is the fastest, we observe some deviation from that baseline due to the contribution of the convective term in $\tau_{\text{SUPG}}$. 
Nevertheless, that variation is not as significant as that of the present formulation where we see large changes in $\tau_{\text{SUPG}}$ depending on the flow velocity, mesh resolution, and flow orientation relative to the element edges. 
 
We also simulated the present flow over a square object case at a higher Reynolds number of $22,000$ to benchmark the two formulations against previously published results\cite{bearman_obasaju_1982,lyn_einav_rodi_park_1995,KOOBUS20041367,rodi1997status}. 
All simulation parameters are kept the same as in the case discussed above except for the dynamic viscosity which is reduced to $2.75 \times 10^{-6}$ kg/m-s. 
Simulations are repeated using both formulations at $\Delta t = 5 \times 10^{-3}$ seconds. 
The obtained results as well as those from the literature are summarized in Table~\ref{tab:2}.
Note that the results from the literature were obtained from three-dimensional simulations whereas our computations are performed in two dimensions.
Despite such differences, we observe a relatively good agreement with the published results. 
That is particularly the case for the present formulation that produces predictions closer to the reported range in comparison to the conventional formulation.

\begin{table}[H]
\centering
\begin{tabular}{l c c c c c} 
 & $\overline{C_\mathrm{d}}$ & $C_\mathrm{d}'$ & $\overline{C_\mathrm{l}}$ & $C_\mathrm{l}'$ & St\\ 
 \hline
 \textbf{Conventional} & 1.62 & 0.40 & $-0.15$ & 1.49 & 0.17 \\
 \textbf{Present} & 1.67 & 0.42 & $-0.03$ & 1.50 & 0.17 \\
 \textbf{Reference}\cite{rodi1997status,bearman_obasaju_1982} & [1.66,2.77] & [0.10, 0.27] & [-0.09 0.03] & [0.34 1.79] & [0.07 0.15] \\
 \hline
\end{tabular}
\caption{\label{tab:2} Results for the flow over a square object at $\mathrm{Re}=22,000$ using the conventional and present formulations, and their comparison against the literature. }
\end{table}

In order to investigate the effect of domain size for the present formulation, we extended the domain so that the overall domain size is four times that of the original domain. 
In doing so, we made sure that the mesh for the subset of the domain corresponding to the original mesh in Figure~\ref{fig:meshsq} remains unchanged so that the reported results are minimally affected by this change in the domain size. 
We repeated the original $\mathrm{Re} = 332$ simulations for both formulations using $\Delta t = 10^{-3}$ with the extended domain.
The results are summarized in Table~\ref{tab:1}.
The variation in results is almost identical for the two formulations indicating that the change is more likely a result of moving the locations of the boundaries rather than the change in the value of $\omega$.
Following what we argued earlier, extending the domain reduced $\omega$ to approximately half its original value.
This change, nevertheless, has a negligible effect on the overall results given that the $\omega$ term is much smaller than the sum of the other terms appearing in $\tau_{\rm SUPG}$.

\begin{table}[H]
\centering
\begin{tabular}{c c c c c c} 
 & $\overline{C_\mathrm{d}}$ & $C_\mathrm{d}'$ & $\overline{C_\mathrm{l}}$ & $C_\mathrm{l}'$ & St\\ 
 \hline
 \multicolumn{1}{l}{\textbf{Conventional}} & & & & & \\
 Original & 1.74 & 0.04 & $-0.00$ & 0.74 & 0.15 \\
 Extended & 1.79 & 0.02 & $-0.00$ & 0.73 & 0.15 \\
 \hline
 \multicolumn{1}{l}{\textbf{Present}} & & & & & \\
 Original & 1.72 & 0.04 & $-0.00$ & 0.83 & 0.15 \\
 Extended & 1.79 & 0.03 & $-0.01$ & 0.82 & 0.15 \\
 \hline
\end{tabular}
\caption{\label{tab:1} Result comparison between the conventional formulation and present formulation using the original and the extended domains for the flow over a square case. $\Delta{t}=10^{-3}$. $\mathrm{Re} = 332$.}
\end{table}

The relatively small value of $\omega$, in comparison to the sum of the two other terms appearing in $\tau_{\rm SUPG}$ in Equation \eqref{eqn:tauom}, is a general feature of the present formulation rather than being unique to the case above. 
Evidently, it was also relatively small for the vascular model discussed earlier as the results obtained from the proposed formulation closely resembled that of the conventional formulation at large $\Delta t$ (Figure~\ref{fig:mbts}), where the sum of the other two terms in $\tau_{\rm SUPG}$ is dominant.

In general, if we only consider $\matr{u}^h \cdot \boldsymbol{\xi} \matr{u}^h$ relative to $\omega^2$, we can show the ratio of the two scales as the square of $u/(\omega \Delta x)$. 
For the present method to be domain-size-dependent, that ratio must be smaller than one, implying that the flow at a node must do a full oscillation before it has the time to advect the fluid across a single element. 
That is an extremely fast oscillating flow, which even if occurs in reality, requires a much smaller time step to resolve, thereby indicating that the present method will do much better than the conventional method.

We should note that the linear solver convergence, which was not an issue for the previous two cases, posed a problem in this case.
In general, the linear solver takes longer to converge as the time step sizes are reduced.
In extreme cases, the linear solver may not converge at all. 
One such scenario occurs when the time step size is very small and there is a significant change in the solution between time steps, e.g., the present simulation starting from a zero velocity field. 
To overcome such issues, one may start a simulation with a larger time step size and then reduce it to the target value after a few time steps. 
Nevertheless, this convergence issue is a shortcoming of the present formulation that warrants future research. 

\subsection*{{Fluid-structure interaction}}
In this section, we will demonstrate the generalization of the present formulation to moving domain configurations using a fluid-structure interaction (FSI) test case. 
The chosen case is a 2D flow over a fixed square with an attached flexible beam, which has been used in the past for verification of the FSI codes\cite{bazilevs2008isogeometric,esmaily2015bi}. 

\begin{figure} [H]
\centering
  \captionsetup{position=bottom}
  \includegraphics[width=0.5\textwidth]{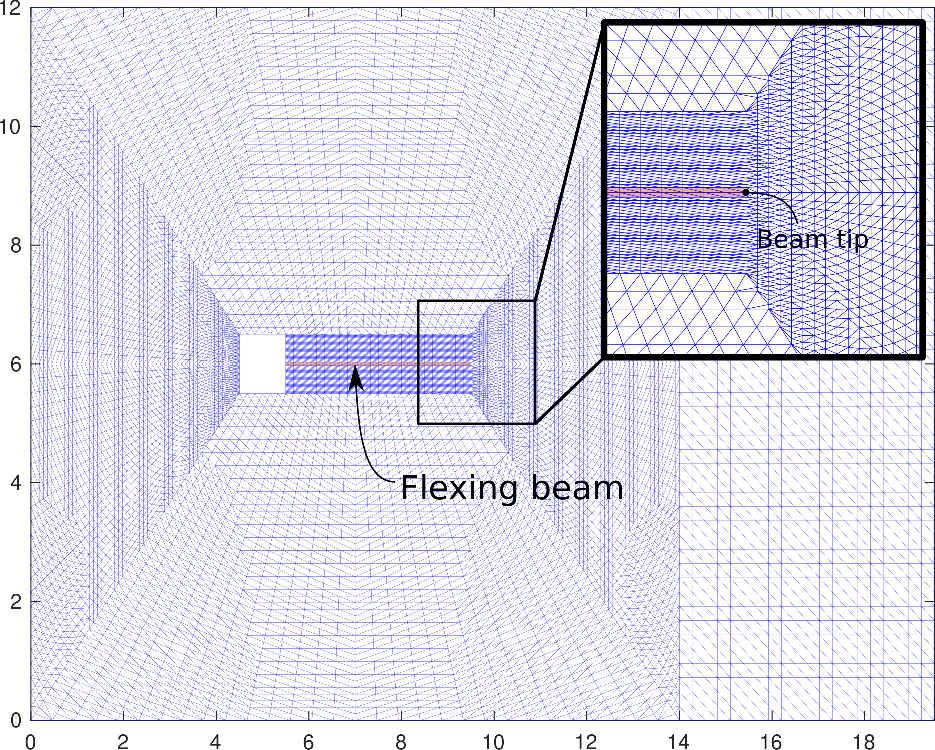}
  \caption{{The schematic of the FSI case involving a flexible beam attached to a solid square in a cross-flow.}}
  \label{fig:flapping}
\end{figure}
The specifications of the test case are shown schematically in Figure~\ref{fig:flapping}. 
The computation domain is $12\times 19.5$ m.
The square is $1\times1$ m with the center of the square placed 5 m from the inlet. 
The beam is $4\times 0.06$ m centered behind the square.
The fluid domain setup is identical to the previous section's flow over a square case, where the fluid density is $1.18 \times 10^{-3}$ kg/m\textsuperscript{3} and the dynamic viscosity is $1.82 \times 10^{-4}$ kg/m-s with uniform inlet horizontal velocity at 51.3 m/s, resulting in Re$=332$. 
The $\omega$ in the stabilization constant, $\tau_{\text{SUPG}}$, is calculated in an arbitrary Eulerian-Lagrangian framework, as specified in Equation~\ref{omega-ale}.

The solid domain is modeled as a St. Venant–Kirchhoff elastic solid\cite{holzapfel2002nonlinear}. 
The discretized weak form of the solid problem is given as
\begin{equation}
    \matr{R}_m^h(\matr{\dot{u}},\matr{d}) = \int_{\Omega_s^0}\left( \rho_s^0 \matr{w}^h \cdot \matr{\dot{u}}^h + \matr{\nabla w}^h:(\matr{F}\matr{\Tilde{S}})^h\right) d\Omega = \matr{0},
    \label{eqn:solid}
\end{equation}
where
\begin{equation}
    \matr{\Tilde{S}} = \lambda \textrm{tr}(\matr{E})\matr{I} + 2\mu\matr{E},
    \label{eqn:PK-stress}
\end{equation}
is the second Piola-Kirchhoff stress, in which
\begin{equation}
    \matr{E} = \frac{1}{2}(\matr{C}-\matr{I})
\end{equation} 
is the Green-Lagrange strain tensor, with 
\begin{equation}
    \matr{C} = \matr{F}^{\text{T}}\matr{F}
\end{equation}
as the Cauchy-Green deformation tensor, where 
\begin{equation}
    \matr{F} = \matr{\nabla d} + \matr{I}
\end{equation}
is the deformation matrix and $\matr{d}$ is the displacement vector.
The density of the beam at $t=0$ is $\rho_s^0 = 0.1 \mathrm{kg/m^3}$.
The Young's modulus, $E = 2.5 \times 10^6$ kg/(s\textsuperscript{2}m), and Poisson's ratio, $\nu = 0.35$ are used to calculate Lam\'{e} parameters $\lambda$ and $\mu$ in Equation~\ref{eqn:PK-stress} as 
\begin{align}
    \mu &= \frac{E}{2(1+\nu)}, \\
    \lambda &= \frac{E\nu}{(1+\nu)(1-2\nu)}.
\end{align}

The details of the FSI formulation are presented in a previous paper and not repeated here for brevity~\cite{esmaily2015bi}. 
In summary, an arbitrary Lagrangian-Eulerian (ALE) approach and a quasi-direct FSI solution strategy are employed\cite{wall1999fluid,bazilevs2008isogeometric,tayfun2007modelling,bazilevs2009computational}. 
The increments of the fluid and structure solution are monolithically computed.
Jacobian-based stiffening is used for the elastic mesh motion without re-meshing\cite{bazilevs2009computational,JOHNSON199473,stein2003mesh}. 
A back-flow stabilization scheme is employed at the Neumann boundaries to prevent simulation divergence caused by partial flow reversal at the outlet\cite{moghadam2011comparison}. 

The mesh used for all cases contains roughly 20 thousand triangular elements for the combined fluid and solid domains (Figure \ref{fig:flapping}). 
The simulations were run for 10 seconds with two different time step sizes $\Delta t = 1\times 10^{-3}$ and $5\times 10^{-4}$ seconds using the conventional and present formulations of $\tau_{\text{SUPG}}$, resulting in four simulations in total. 

\begin{figure} [H]
\centering
  \captionsetup{position=bottom}
  \includegraphics[width=\textwidth]{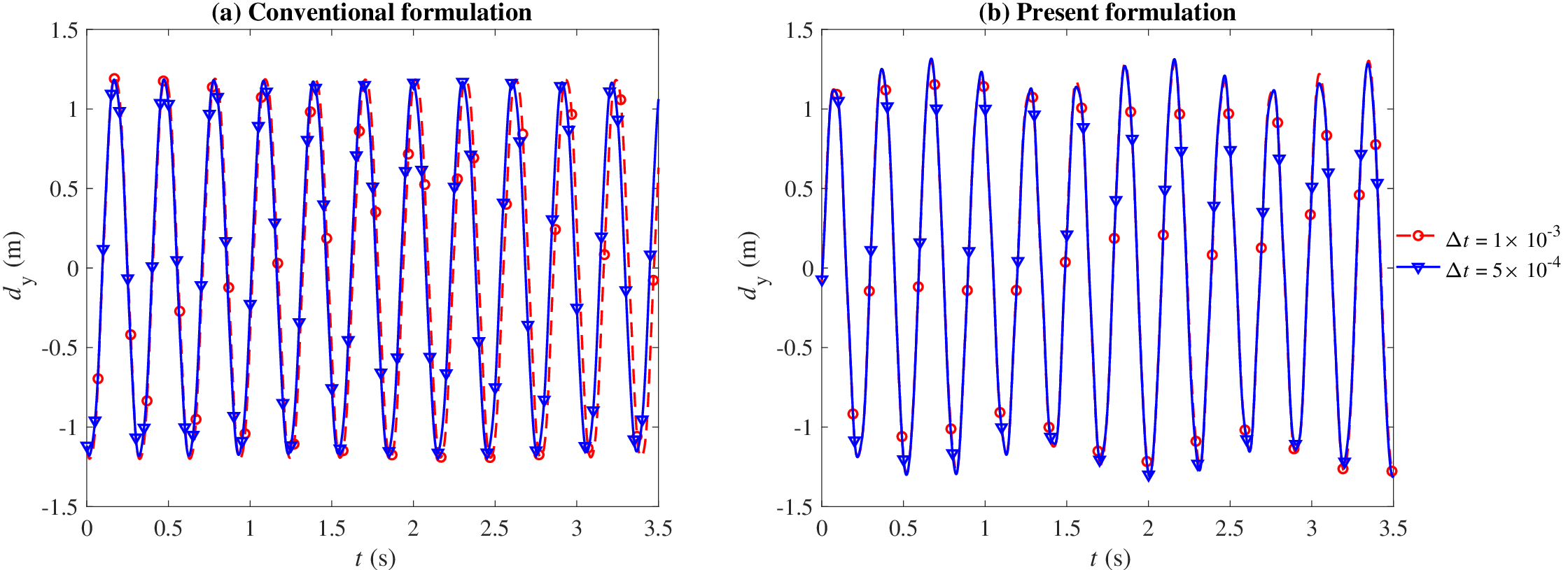}
  \caption{{The beam tip vertical displacement, $d_\mathrm{y}$, for the case shown in Figure \ref{fig:flapping} computed using the conventional (a; left) and present formulation (b; right). The two curves shown as a function of time for each case correspond to the simulations run with $\Delta t = 1\times 10^{-3}$ (dashed line) and $5\times 10^{-4}$ (solid line).}}
  \label{fig:fsi-dy}
\end{figure}

Figure~\ref{fig:fsi-dy} shows the vertical displacement of the beam tip (marked in Figure \ref{fig:flapping}). 
The conventional formulation produces an oscillation period that slightly varies as the time step size is changes by a factor of two.
The change in period is approximately 0.7\%, increasing from 0.305 s at $\Delta t = 5\times 10^{-4}$ s to 0.307 s at $\Delta t = 10^{-3}$ s.
On the contrary, the present formulation produced significantly closer results for both time step sizes. 
The predicted period, in this case, is approximately 0.299 s, which agrees well with the literature \cite{wall1999fluid,bazilevs2008isogeometric,esmaily2015bi}. 

In terms of stability and computational cost, we observe a behavior similar to the flow over the square case from the previous section. 
For cases reported above, the total simulation cost of the present formulation is around 4 times that of the conventional formulation (1.3 versus 5.1 hours using 16 cores).
Using a time step size smaller than $\Delta t = 5\times 10^{-4}$ s causes linear solver convergence issues in the case of the present formulation.

\section*{Conclusion}
\label{sec:con}
In this study, we propose a new formulation for the stabilization parameter ($\tau_{\text{SUPG}}$) that appears in the streamline upwind Petrov-Galerkin and pressure stabilizing Petrov-Galerkin method (SUPG/PSPG) to overcome the historical limitation of the conventional formulation that produces a large error at the small time step size.
The proposed formulation uses a flow time scale instead of the time step size to account for the contribution of the acceleration term to $\tau_{\text{SUPG}}$.
Using the new formulation of $\tau_{\text{SUPG}}$, we successfully produce an overall stable technique that is consistent with regard to the time step size. 
The new definition of $\tau_{\text{SUPG}}$ (Equation \eqref{eqn:tauom}) is simple to implement in existing SUPG/PSPG formulated fluid solvers. 
Although the present formulation comes at the cost of increasing the number of linear solver iterations, it significantly improves the overall accuracy of the stabilized finite element methods for computational fluid dynamics, making it an attractive choice for cardiovascular simulations.

\section*{Data availability}
The software code to the finite element solver used in this study (MUPFES) is available at \url{sites.google.com/site/memt63/tools/MUPFES}.
The simulation files and results presented in this study are available on request from the corresponding author.

\bibliography{ref}

\section*{Acknowledgements}

We thank Prof. John Evans for bringing our attention to the existing literature on the topic and for his insights on the stability properties of the proposed approach. We also thank Prof. Yuri Bazilevs for his input during the early stages of this study that led to the selection of the flow over square vortex shedding test case. 

\section*{Author contributions statement}
Conceptualization, M.E.; methodology, D.J. and M.E.; software, D.J. and M.E.; validation, D.J. and M.E.; formal analysis, D.J.; investigation, D.J.; resources, M.E.; data curation, D.J. and M.E.; writing---original draft preparation, D.J.; writing---review and editing, D.J. and M.E.; visualization, D.J.; supervision, M.E.; project administration, M.E.. All authors have read and agreed to the published version of the manuscript.

\section*{Additional Information}
\subsection*{Competing interest statement}
The authors declare no financial or non-financial competing interests.

\end{document}